\documentclass[a4paper]{article}
\usepackage[utf8]{inputenc}
\usepackage[T1]{polski}
\usepackage{graphicx}\graphicspath{{img/}}
\usepackage{lmodern,fancyhdr,secdot,float,amsmath,amsfonts,amsthm,amssymb}
\usepackage{booktabs,tabularx}
\usepackage[margin=2cm]{geometry}
\usepackage{multicol}
\usepackage[hidelinks,unicode]{hyperref}
\usepackage{titlesec}
\titleformat{\section}{\centering\normalsize\bf}{\thesection.}{.5em}{\MakeUppercase}
\titleformat*{\subsection}{\bf\normalsize\selectfont}
\titleformat*{\subsubsection}{\bf\normalsize\selectfont}
\newcommand{\titlePL}[1]{\large\textbf{ #1}}
\newcommand{\titleEN}[1]{\normalsize #1}
\newcommand{\keywordsPL}[1]{\small\textbf{Słowa kluczowe:} #1}
\newcommand{\keywordsEN}[1]{\small\textbf{Keywords:} #1}
\newcommand{\abstractPL}[1]{\small\textbf{Streszczenie:} #1}
\newcommand{\abstractEN}[1]{\small\textbf{Abstract:} #1}

\begin{document}\thispagestyle{empty}\pagestyle{fancy}
\begin{minipage}[t]{0.5\textwidth}\vspace{0pt}%
\includegraphics[scale=1.1]{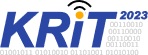}
\end{minipage}
\begin{minipage}[t]{0.45\textwidth}\vspace{0pt}%
\centering
KONFERENCJA RADIOKOMUNIKACJI\\ I TELEINFORMATYKI\\ KRiT 2023
\end{minipage}

\vspace{1cm}

\begin{center}
\titlePL{WYKORZYSTANIE REKONFIGUROWALNYCH MATRYC ANTENOWYCH WRAZ Z INFORMACJĄ KONTEKSTOWĄ}

\titleEN{UTILIZATION OF RECONFIGURABLE INTELLIGENT SURFACES WITH CONTEXT INFORMATION}\medskip

Łukasz Kułacz$^{1}$,
Adrian Kliks$^{2}$

\medskip

\begin{minipage}[t]{0.6\textwidth} 
\small $^{1}$ Politechnika Poznańska, Poznań, \href{mailto:email}{lukasz.kulacz@put.poznan.pl} \\
\small $^{2}$ Politechnika Poznańska, Poznań, \href{mailto:email}{adrian.kliks@put.poznan.pl}
\end{minipage}

\medskip

\end{center}

\medskip

\begin{multicols}{2}
\noindent
\abstractPL{ 
Rekonfigurowale matryce antenowe mogą zostać użyte z powodzeniem do sterowania środowiskiem radiowym. Proste sterowanie kątem odbicia sygnału od powierzchni pozwala na maksymalizację lub minimalizację odbieranej mocy w określonych miejscach. W~artykule przedstawiono symulacje, gdzie możliwy jest odbiór sygnału w miejscu, w~którym nie było to możliwe, wykrycie zajętości widma w miejscu, w którym detektor nie był w~stanie dokonać prawidłowej detekcji, czy zminimalizowanie zakłóceń w~konkretnym odbiorniku. }
\medskip

\noindent
\abstractEN{ 
Reconfigurable intelligent surfaces can be successfully used to control the radio environment. Simple control of the reflection angle of the signal from the surface allows maximization or minimization of the received power in specific places. The paper presents simulations where it is possible to receive a signal in a place where it was not possible, to detect the occupancy of the spectrum in a place where the sensor was unable to make correct detection or to minimize interference in a specific receiver.}
\medskip

\noindent
\keywordsPL{detekcja zajętości widma, informacja kontekstowa, propagacja sygnału, rekonfigurowalne matryce antenowe}
\medskip

\noindent
\keywordsEN{context information, reconfigurable intelligent surfaces, signal propagation, spectrum occupancy detection.}

\section{Wstęp}
Rozwój komunikacji bezprzewodowej w ostatnich czasach opiera się w bardzo często na wprowadzaniu do sieci coraz większego poziomu inteligencji. Pod pojęciem wspomnianej inteligencji uznaje się między innymi systemy dynamicznego dostępu do widma, które często wymagają dostępu do dużej ilości informacji kontekstowej o środowisku i użytkownikach sieci~\cite{b4}. Informacje te są niezbędne do podejmowania optymalnych decyzji – w oparciu o~wiele parametrów (niekoniecznie związanych bezpośrednio z samą siecią) oraz ich historyczne wartości. Dane historyczne pozwalają między innymi wychwycić trendy w~zachowaniach użytkowników sieci, a tym samym umożliwiają przewidywanie nadchodzących wydarzeń. Środowisko miejskie jest niezwykle wymagające dla projektantów systemów bezprzewodowych, gdzie przesyłany sygnał poddawany jest wielu zjawiskom propagacyjnym, tj. wielokrotnym odbiciom czy rozproszeniom. Ogromna liczba i~różnorodność budynków i obiektów wpływających łącznie na propagację sygnału tworzy bardzo skomplikowany kanał radiowy. Dodatkowo w środowisku miejskim występuje ogromna liczba użytkowników na stosunkowo niewielkim obszarze. Połączenie tych dwóch faktów stanowi spore wyzwanie w~kontekście działania systemów bezprzewodowych.
Warto zauważyć, że generalnie odbicia sygnału są nieuniknione i choć utrudniają analizę propagacji sygnału, to można uznać je za zjawisko korzystne - ponieważ sygnał ma możliwość dotarcia do odbiornika, który nie znajduje się bezpośrednio w polu widzenia nadajnika. W tym kontekście szczególnie ciekawym podejściem jest popularny i rozważany w ostatnim czasie temat Rekonfigurowalnych Matryc Antenowych, RMA (ang. Reconfigurable Intelligent Surfaces, RIS)~\cite{b1}. RMA pozwalają wpływać na proces odbicia fal radiowych na swojej powierzchni. Oznacza to, że możliwa jest modyfikacja podstawowej zasady, w której kąt padania fali jest równy kątowi odbicia fali. Stwarza to ogromne możliwości w kreowaniu i~sterowaniu środowiskiem propagacyjnym. Z~jednej strony kierowanie sygnału radiowego w miejsca wcześniej niedostępne (ze względu na geometrię obiektów). Z drugiej strony usunięcie sygnału radiowego w określonym miejscu (minimalizacja zakłóceń) poprzez skierowanie go w zupełnie innym kierunku~\cite{b2}. Tylko te dwie koncepcje wydają się mieć duży potencjał do wykorzystania w sieciach inteligentnych i~mogą służyć do wykrywania sygnału lub poprawy jakości odbieranego sygnału (poprzez zwiększenie odbieranej mocy, jeśli jest to pożądany sygnał lub zmniejszenie odbieranej mocy, jeśli jest to niepożądany sygnał). Wykrywanie sygnału mogłoby być możliwe dla detektora, który jest w~jakiś sposób ukryty przed nadajnikiem nie tylko w~sensie bezpośredniej widoczności, ale także sygnałów odbitych o odpowiednio dużej mocy niezbędnej do działania algorytmów detekcji sygnału. Wskazane rozwiązania wymagają jednak opracowania nowych algorytmów, których zadaniem jest sterowanie opisaną RMA~\cite{b3}.

\section{Opis analizowanego scenariusza}
W tym artykule rozważane jest wydłużone pomieszczenie, w którym między nadajnikiem sygnału a trzema odbiornikami znajduje się ściana uniemożliwiająca bezpośrednią widoczność urządzeń. W głębi pomieszczenia znajduje się jednak powierzchnia, która będzie symulować działanie RMA.

\begin{figure}[H]
\centering
\includegraphics[width=0.45\textwidth]{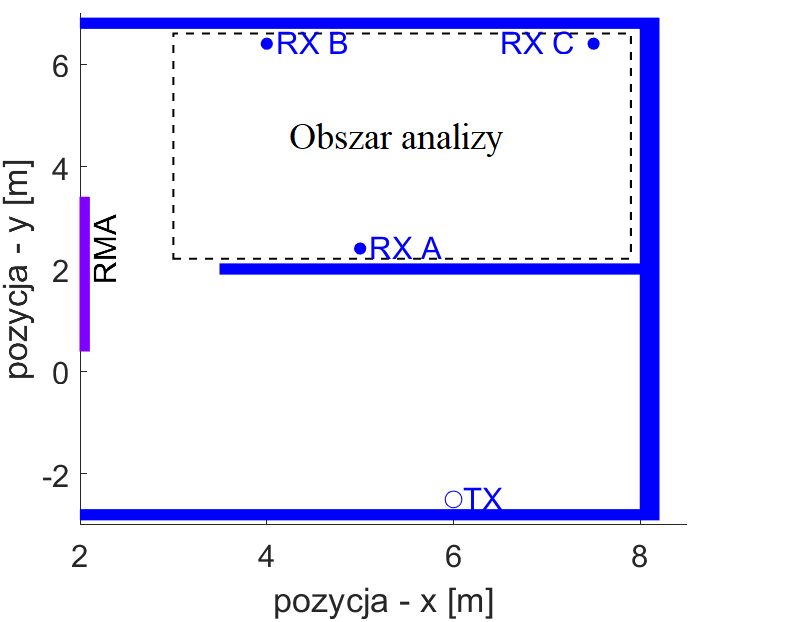}
\caption{Obszar analizy}
\label{rys1}
\end{figure}

Wszystkie urządzenia znajdują się na tej samej wysokości, a wpływ sufitu i podłogi został pominięty – upraszczając analizowany przypadek do przestrzeni dwuwymiarowej. Rzut analizowanego obszaru przedstawiono na Rys.~\ref{rys1}.
Opisywane pomieszczenie zostało wybrane celowo jako demonstracja potencjalnych możliwości wykorzystania RMA. Decyzja ta podyktowana jest chęcią naświetlenia koncepcji wykorzystania RMA, a same pomysły mogą być wykorzystane w bardziej złożonych analizach, gdzie jednak znacznie trudniej byłoby wyizolować wpływ RMA (dla celu demonstracji). W artykule porównano trzy scenariusze symulacji:
\begin{itemize}

\item \textit{Scenariusz 1} --- zakłada brak użycia RMA, dlatego w jego miejscu umieszczony został fragment ściany. Ten scenariusz stanowi odniesienie dla pozostałych scenariuszy.

\item \textit{Scenariusz 2} --- zakłada uproszczone działanie RMA, które jest niezależne od czynników zewnętrznych, tj. okresowo i naprzemiennie zmieniany jest kąt odbitego sygnału. Symuluje to użycie RMA jako płaszczyzny odbijającej (analogicznie do lustra, tylko zakresie fal elektromagnetycznych). Scenariusz ten został zaproponowany w celu wykazania potencjalnych korzyści płynących z zastosowania nawet tak prostego mechanizmu sterowania.

\item \textit{Scenariusz 3} --- to sytuacja, w której użytkownicy sieci mają dostęp do informacji o bieżącym ustawieniu RMA i mają pośredni wpływ na to ustawienie. Ostatni scenariusz to przykład wykorzystania informacji kontekstowej (w tym przypadku aktualnego ustawienia RMA) do osiągnięcia bieżącego celu (np. minimalizacji zakłóceń).
  
\end{itemize}

W artykule przeanalizowano trzy przypadki użycia RMA sprawdzając wszystkie trzy scenariusze (w każdym przypadku użycia). W szczególności rozważono poprawę skuteczności wykrywania zajętości widma przez detektor; zwiększenie mocy sygnału odbieranego w odbiorniku; redukcja zakłóceń w odbiorniku. Podczas symulacji symulowano działanie RMA jako zmianę kąta obrotu, pod jakim  zaznaczona na Rys.~\ref{rys1} powierzchnia (jako RMA) znajduje się względem równoległej do niej ściany (po prawiej stronie rysunku). Rozważono wartości kąta od -20 do 20 stopni z krokiem 5 stopni.

\section{Wyniki symulacji}

\begin{figure}[H]
\centering
\includegraphics[width=0.45\textwidth]{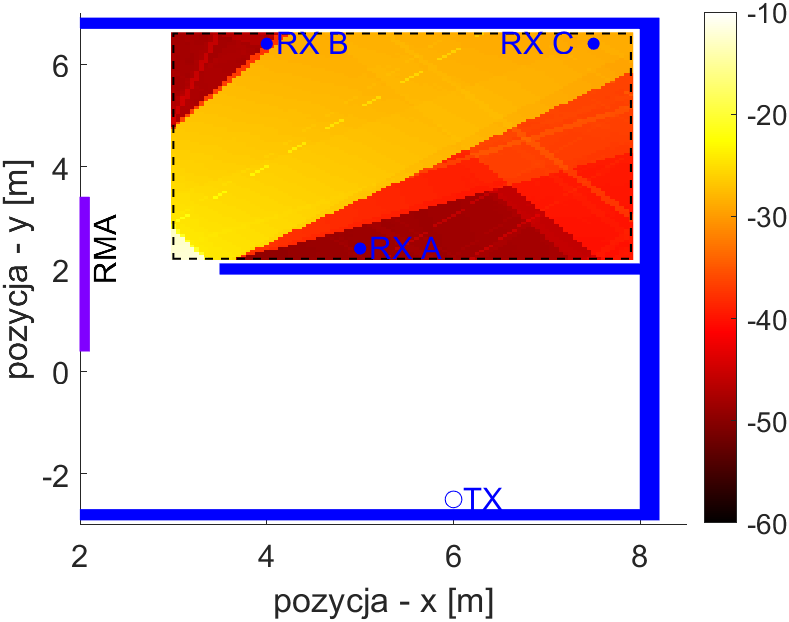}
\caption{Odbierana moc sygnału bez RMA [dBm] (scenariusz 1)}
\label{rys2}
\end{figure}

Na Rys.~\ref{rys2} przedstawiono rozkład przestrzenny mocy odbieranej na analizowanym obszarze dla zadanych parametrów transmisyjnych i przy wykorzystaniu modelu propagacji sygnału opartego na śledzeniu promieni. Do przeprowadzenia symulacji wykorzystane zostało oprogramowanie MATLAB (z zestawem narzędzi ''Antenna Toolbox'').
Ze wstępnej analizy przedstawionego rozkładu wynika, że część obszaru charakteryzuje się mniejszym tłumieniem sygnału (barwy cieplejsze, odbiornik C), natomiast pozostała część obszaru ma większe tłumienie sygnału (kolory ciemniejsze, odbiorniki A i B – pozostające ''w cieniu'' sygnału). Moc odbierana w tej sytuacji wynika bezpośrednio z prostopadłego ułożenia ściany oznaczonej na Rys.~\ref{rys1} jako ''RMA'' – tu jednak pełniącej rolę fragmentu ściany. Można zauważyć, że odbiorniki A i B miałyby problem zarówno z odbiorem, jak i wykryciem obserwowanego sygnału, byłyby jednak pozbawione nadmiernych zakłóceń. Sytuacja jest dokładnie odwrotna dla odbiornika C. Należy jednak pamiętać, że w odwrotnej sytuacji, gdy odbiorniki A i B uznają analizowany sygnał za niepożądany, a odbiornik C za sygnał pożądany, to oddziaływanie na propagację nie byłoby korzystne. 
\begin{figure}[H]
\centering
\includegraphics[width=0.45\textwidth]{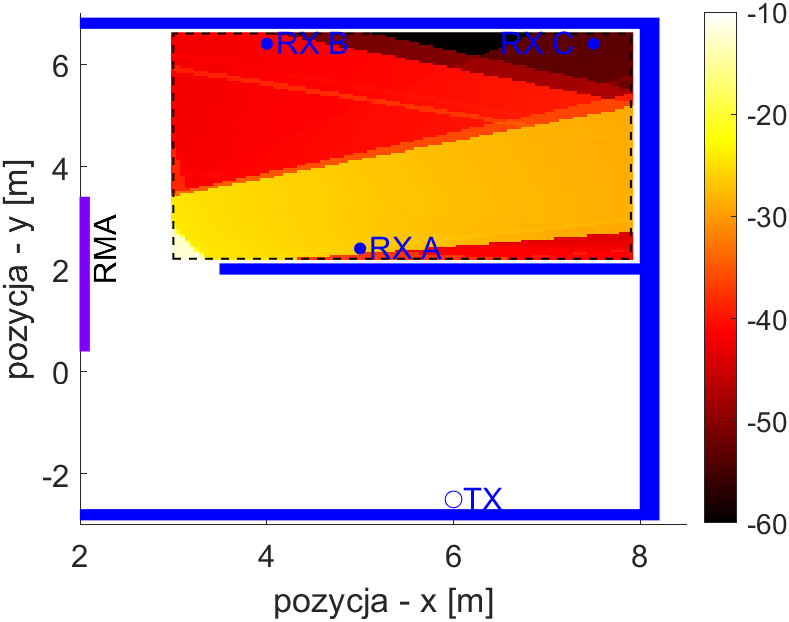}
\caption{Odbierana moc sygnału z obecnym RMA [dBm] (scenariusz 3)}
\label{rys3}
\end{figure}
Dla porównania na Rys.~\ref{rys3} przedstawiony został rozkład przestrzenny mocy odbieranej dla sytuacji, gdy RMA jest w użyciu i obija sygnał w taki sposób, że odpowiada to obróceniu fragmentu ściany o 20 stopni. Analiza tego rozkładu ukazuje, że odbiornik A, obserwuje w tej sytuacji o wiele większą moc odbieraną w stosunku do poprzedniej analizy (przedstawionej na Rys.~\ref{rys1}). W przeciwnej sytuacji znajduje się odbiornik B. W następnych podrozdziałach przeanalizowane zostaną kolejne przypadki użycia RMA w symulowanym środowisku.

\subsection{Pierwszy przypadek użycia: wykrywanie zajętości widma}

\begin{figure}[H]
\centering
\includegraphics[width=0.45\textwidth]{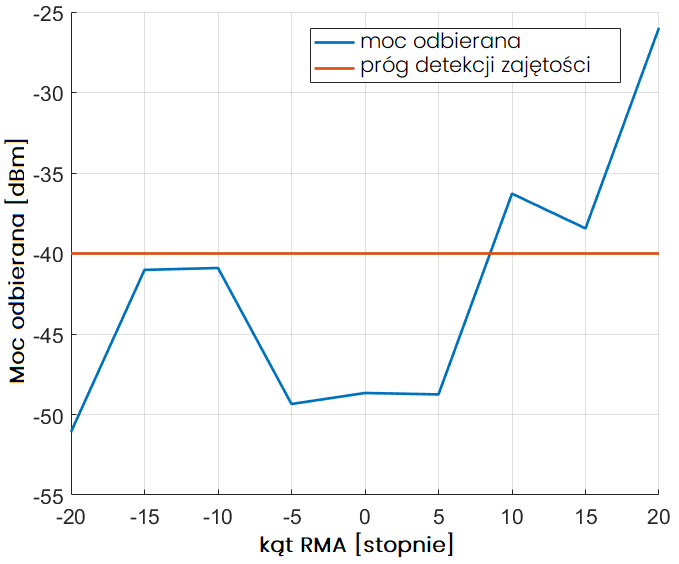}
\caption{Odbierana moc w odbiorniku A (scenariusz 2)}
\label{rys4}
\end{figure}

Dla pierwszego przypadku użycia tj. poprawy skuteczności detekcji zajętości widma przez odbiornik A, symulacje wykazały, że w pierwszym scenariuszu poprawna detekcja jest bardzo utrudniona (wręcz niemożliwa), w drugim scenariuszu poprawna detekcja jest możliwa przez około 31,25\% czasu symulacji, a w trzecim scenariuszu, przez około 79,17\% czasu istnieje możliwość poprawnej detekcji. Moc obserwowana w odbiorniku A w drugim scenariuszu (w zależności od ustawienia kąta RMA) przedstawiono na Rys.~\ref{rys4}. 

Oczywiście ostatni scenariusz jest silnie uzależniony od zastosowanego algorytmu (i jego przeznaczenia). Jednakże nawet informacja o aktualnym ustawieniu RMA jest wystarczająca, aby detektor aktualizował decyzje o~zajętości widma tylko dla odpowiednich ustawień RMA. To podejście nie jest idealne, a jego skuteczność zależy w~dużej mierze od charakterystyki transmisji (częstotliwość, długość transmisji itp.). Rozwiązanie zakładające możliwość sterowania RMA nawet w bardzo prostym scenariuszu umożliwia np. okresowe ''przeczesywanie'' wszystkich dostępnych ustawień, a następnie utrzymywanie jednego ustawienia (lub kilku na przemian) przez określony czas. W rozważanym systemie z jednym detektorem po krótkim czasie obserwacji możliwe jest ustawienie RMA na cały pozostały czas symulacji, co pozwala osiągnąć możliwość prawidłowej detekcji przez prawie 100\% czasu.

\subsection{Drugi przypadek użycia: Poprawa poziomu odbieranego sygnału}

\begin{figure}[H]
\centering
\includegraphics[width=0.45\textwidth]{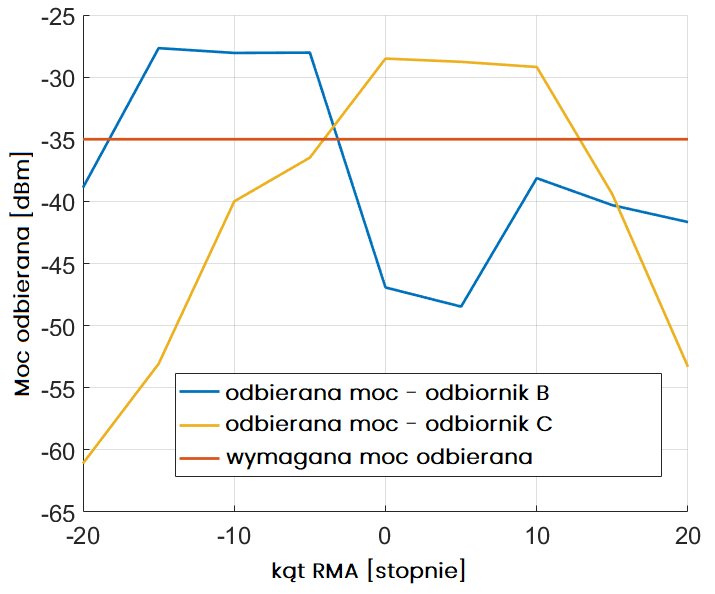}
\caption{Odbierana moc w odbiornikach B i C }
\label{rys5}
\end{figure}

Dla drugiego przypadku użycia, tj. poprawy poziomu odbieranego sygnału dla odbiorników B i C, wyniki symulacji pokazują, że w pierwszym scenariuszu odbiornik B nie jest w stanie poprawnie odebrać sygnału, podczas gdy odbiornik C może odbierać sygnał poprawnie przez cały czas. W drugim scenariuszu, gdzie ustawienie RMA jest zmieniane okresowo, przez 35,29\% czasu odbiornik B odbiera sygnał o wystarczającej mocy, podczas gdy odbiornik C przez 41,18\% czasu. Moc odbierana w zależności od ustawionego kąta RMA dla scenariusza drugiego została przedstawiona na Rys.~\ref{rys5}. W tym przypadku widzimy zysk w postaci możliwości obsługi odbiornika, który znajdował się w ''cieniu'' sygnału i nie był w~stanie odebrać sygnału bez zmiany swojego położenia. W trzecim scenariuszu przyjęto, że przez początkowy czas RMA sprawdza wszystkie dostępne ustawienia, a następnie na podstawie zgłoszeń od odbiorników naprzemiennie korzysta ze zgłoszonych (najlepszych dla poszczególnych odbiorników) ustawień. W rezultacie odbiornik B był w~stanie odbierać sygnał przez 45,1\% czasu, a odbiornik C przez 47,06\% czasu. W tym miejscu warto wspomnieć o możliwości uwzględnienia algorytmów sterowania ruchem w~celu dostosowania czasu trwania ustawień dla poszczególnych odbiorników.

\subsection{Trzeci przypadek użycia: redukcja zakłóceń}

Dla trzeciego przypadku użycia tj. redukcji zakłóceń w~odbiorniku C, założono dopuszczalny próg mocy sygnału zakłócającego, powyżej którego odbiornik jest nadmiernie zakłócany. Wyniki symulacji pokazują, że w~pierwszym scenariuszu przez 100\% czasu odbiornik obserwuje znaczne zakłócenia (powyżej założonego progu). W~drugim scenariuszu zakłócenia przekraczają próg przez 35,29\% czasu. Warto w tym miejscu wspomnieć, że średni obserwowany poziom zakłóceń spada o około 3,715 dB, mediana o około 7,972 dB, a 10 percentyl o około 24,762 dB. Oczywiście 90-ty percentyl pozostaje taki sam. Znajomość ustawienia RMA mogłaby przynajmniej dostarczyć informacji o momencie wystąpienia potencjalnych zakłóceń – co można np. wykorzystać w procesie planowania transmisji. W~trzecim scenariuszu ponownie zakładamy określony czas na sprawdzenie wszystkich ustawień RMA, a następnie na podstawie raportu odbiornika C ustawienie konfiguracji powodującej minimalny poziom zakłóceń. Dzięki temu udało się skrócić czas zakłóceń powyżej założonego progu z 100\% czasu do 19,61\% czasu. Wyznaczony czas ponownie jest mocno zależny od danej konfiguracji i obecności innych urządzeń.

Należy pamiętać, że przeprowadzone symulacje zakładały stały w czasie schemat transmisji podczas całego eksperymentu. W przypadku jedynie chwilowych transmisji sygnału dla wszystkich trzech przypadków użycia byłoby to dodatkowe wyzwanie dla algorytmu sterującego pracą RMA.

\subsection{Wszystkie przypadki użycia łącznie}

Dodatkowo przeprowadzono eksperyment symulacyjny łączący wszystkie trzy przedstawione przypadki użycia, tj. odbiornik A jest detektorem wykrywającym zajętość widma, odbiornik B jest odbiornikiem analizowanego sygnału (sygnału pożądanego), a odbiornik C nie jest odbiornikiem analizowanego sygnału (sygnał niepożądany). Zauważmy, że w scenariuszu referencyjnym cele żadnego z odbiorników nie są spełnione. Następnie zastosowano RMA sterowany według następującego schematu: początkowo sprawdzane są wszystkie możliwe ustawienia kąta RMA, a następnie naprzemiennie wykorzystywane są wszystkie najlepsze ustawienia zgłoszone przez odbiorniki. W wyniku symulacji odbiornik A jest w stanie z powodzeniem wykryć sygnał przez 32,08\% czasu, odbiornik B jest w stanie z powodzeniem odebrać pożądany sygnał przez 33,96\% czasu, a odbiornik C jest wolny od znaczących zakłóceń przez 79,25\% czasu. Dodatkowo średnia moc zakłóceń obserwowana w odbiorniku C spada o~około 8,6 dB. Dodatkowo sprawdzone zostało jeszcze jedno rozwiązanie, gdzie po sprawdzeniu wszystkich ustawień kąta RMA wybrane zostają na resztę symulacja dwa (zamiast trzech) ustawienia, dla których nadal spełnione są założone wymagania. W efekcie wybierane są na przemian dwa ustawienia: najlepsze dla detektora A oraz najlepsze dla odbiornika B. Decyzja ta została uzasadniona minimalizacją liczby rozwiązań, a tym samym liczby zmian w systemie, oraz tym, że najlepsze ustawienie dla odbiornika B jest tylko nieco gorszym rozwiązaniem dla odbiornika C. Uzyskano w ten sposób następujące rezultaty. Odbiornik A jest w stanie z powodzeniem wykryć sygnał przez 43,4\% czasu, odbiornik B jest w stanie z powodzeniem odebrać pożądany sygnał przez 45,28\% czasu, a odbiornik C jest wolny od znaczących zakłóceń przez 79,25\% czasu. Średnia moc zakłóceń obserwowana w odbiorniku C spada o~około 8,58 dB. Zatem udało się tym prostym zabiegiem znacznie poprawić jakoś obsługi odbiorników A i B, kosztem podniesienia średniej interferencji w odbiorniku C o 0,02 dB (średnia obserwowana interferencja nadal poniżej ustalonego dopuszczalnego progu).

\section{Podsumowanie}
Przedstawione w artykule wyniki symulacji pokazują, że wykorzystanie RMA, nawet przy najprostszym mechanizmie sterowania, może przynieść korzyści w każdym z analizowanych przypadków użycia. Jednakże znacznie lepsze efekty można uzyskać implementując algorytm sterowania RMA i mechanizm dostarczania zainteresowanym odbiorcom informacji o aktualnym ustawieniu RMA. Daje to bardzo elastyczne rozwiązanie, którego działanie wymaga dalszej analizy. W szczególności należy wziąć pod uwagę charakter otoczenia, charakter źródła lub źródeł sygnału, potencjalne odbiorniki sygnału, ich położenie oraz cel użytkowania. Przedstawione koncepcje przede wszystkim stwarzają możliwość kierowania sygnału w miejsca wcześniej niedostępne – jednak należy pamiętać, że sytuacja ta może być zarówno pozytywna, jak i negatywna. Istnieje bowiem ryzyko wprowadzenia zakłóceń w miejscu, w którym dotychczas nie było nadmiernych zakłóceń. Problem ten pogłębia się zwłaszcza w odniesieniu do urządzeń nieświadomych istnienia i obecności RMA. Niemniej jednak pomysły przedstawione w tym artykule wydają się obiecujące pod względem potencjalnych korzyści dla pewnych zastosowań.

\section*{Podziękowanie}
Praca powstała w ramach projektu finansowanego przez Narodowe Centrum Nauki (nr 2021/43/B/ST7/01365).


\end{multicols}
\end{document}